\documentclass[conference]{IEEEtran}
\makeatletter
\def\ps@headings{%
\def\@oddhead{\mbox{}\scriptsize\rightmark \hfil \thepage}%
\def\@evenhead{\scriptsize\thepage \hfil \leftmark\mbox{}}%
\def\@oddfoot{}%
\def\@evenfoot{}}
\makeatother
\ifCLASSINFOpdf
\else
\fi
\usepackage{cite}
\usepackage{amssymb}
\usepackage{subfigure}
\usepackage{graphicx}
\usepackage[ruled,vlined]{algorithm2e}
\usepackage{float}
\usepackage[T1]{fontenc}
\usepackage{url}
\newtheorem{theorem}{Theorem}[section]

\newenvironment{proof}[1][Proof]{\begin{trivlist}
\item[\hskip \labelsep {\bfseries #1}]}{\end{trivlist}}

\renewcommand {\IEEEQED}{\IEEEQEDopen}

\begin{document}
\title{DiFS: Distributed Flow Scheduling for Data Center Networks\vspace{-1ex}}
\author{\IEEEauthorblockN{Wenzhi Cui}
\IEEEauthorblockA{Software Institute,
Nanjing University, China\\
Email: cwz10@software.nju.edu.cn}
\and
\IEEEauthorblockN{Chen Qian}
\IEEEauthorblockA{Department of Computer Science,
UT Austin, USA\\
Email: cqian@cs.utexas.edu}}

\maketitle

\begin{abstract}
Data center networks leverage multiple parallel paths connecting end host pairs to offer high bisection bandwidth for cluster computing applications. However, state of the art distributed multi-pathing protocols such as Equal Cost Multipath (ECMP) use static flow-to-link assignment, which is load-oblivious. They may cause bandwidth loss due to \emph{flow collisions} on a same link. Recently proposed centralized scheduling algorithm or host-based multi-pathing may suffer from scalability problems.

In this paper, we present Distributed Flow Scheduling (DiFS) for data center networks, which is a \emph{switch-only distributed solution}. DiFS allows switches cooperate to avoid over-utilized links and find available paths without centralized control. DiFS is scalable and can react quickly to dynamic traffic, because it is independently executed on switches and requires no synchronization.  Extensive experiments show that the aggregate bisection bandwidth of DiFS using various traffic patterns is much better than that of ECMP, and is similar to or higher than that of a recent proposed centralized scheduling algorithm.
\end{abstract}

\section{Introduction}
The growing importance of cloud-based applications and big data processing has led to the deployment of large-scale data center networks that carry tremendous amount of traffic. Recently proposed data center network architectures primarily focus on using commodity Ethernet switches to build multi-rooted trees such as fat-tree \cite{Al-Fares08} and Clos \cite{VL2}. These topologies provide multiple equal-cost paths for any pair of end hosts and hence significantly increase bisection bandwidth. To fully utilize the path diversity, an ideal routing protocol should allow flows to avoid over-utilized links and take alternative paths.

Most state of the art data center networks rely on layer-3 Equal Cost Multipath (ECMP)  protocol \cite{ECMP} to assign flows to available links using static flow hashing. Being simple and efficient, however, ECMP is load-oblivious, because the flow-to-path assignment does not account current network utilization. As a result, ECMP may cause flow collisions on particular links and create hot spots.

Recently proposed methods to improve the bandwidth utilization in data center networks can be classified in two categories: \emph{centralized and distributed solutions}. A typical centralized solution Hedera \cite{Hedera} relies on a central controller to find a path for each flow or assign a single core switch to deal with all flows to each destination host. Centralized solutions may face scalability problems, because traffic in today's data center networks requires parallel and fast path selection according to recent measurement studies \cite{IMC09, Benson10}.

The first type of distributed solutions is \emph{host-based}, such as multipath TCP (MPTCP) \cite{MPTCP} and DARD \cite{DARD}. These methods enable end systems to select multiple paths for flows based on network conditions. However, all legacy systems and applications running these protocols need to be upgraded, which incurs a lot management cost. In addition, every host running DARD needs to monitor the states of all paths to other hosts. For many applications such as Shuffle (described in Section \ref{sec:shuffle}), each DARD host may have to monitor the entire network, which also limits its scalability. The second type is \emph{switch-only protocols} which is fully compatible to current systems and applications on end hosts. Many existing switch-only protocols allow a flow take multiple paths at the same time (called flow splitting) to achieve high throughput \cite{Dixit2011, LocalFlow, Zahavi2012}. Flow splitting may cause a high level of TCP packet reordering, resulting throughput drop \cite{reorder}.

In this paper, we propose Distributed Flow Scheduling (DiFS), a switch-only protocol that is executed independently on the control unit of each switch. DiFS aims to balance flows among different links and improves bandwidth utilization for data center networks. DiFS does not need centralized control or changes on end hosts. In addition DiFS does not allow flow splitting and hence limits packet reordering.

DiFS achieves load balancing by taking efforts in two directions. First, each switch uses the \emph{Path Allocation} algorithm that assigns flows evenly to all outgoing links to avoid \emph{local flow collisions}. Second, each switch also  monitors its incoming links by running the \emph{Imbalance Detection} algorithm. If a collision is detected, the switch will send an Explicit Adaption Request (EAR) message that suggests the sending switch of a flow to change its path. Upon receiving the EAR, the sending switch will run the \emph{Explicit Adaption} algorithm to avoid \emph{remote flow collisions}.

We have implemented DiFS on a stand-alone simulator which \emph{simulates behaviors of every packet}. TCP New Reno is implemented in detail as the transportation layer protocol. This simulator is more detailed than those of other flow scheduling papers that only simulate flow behaviors, such as that in \cite{Hedera}. Experimental results show that DiFS outperforms ECMP significantly in aggregate bisection bandwidth. Compared with a centralized solution Hedera, DiFS achieves comparable or even higher aggregate bisection bandwidth, higher throughput, and less out-of-order packets, for both small and large data center network topologies.

The rest of this paper is organized as follows. Section \ref{sec:background} introduces background knowledge of flow scheduling in data center networks. Section \ref{sec:design} presents the detailed architecture and algorithm design of DiFS. We evaluate the performance of DiFS and compare it with other solutions in Section \ref{sec:simulation}. We conclude our work in Section \ref{sec:conclusion}.

\section{Background and Overview of DiFS}
\label{sec:background}
\subsection{Data center topologies}
Today's data center networks often use \emph{multi-rooted tree} topologies (e.g., fat-tree \cite{Al-Fares08} and Clos \cite{VL2} topologies) to exploit multiple parallel paths between any pair of host to enhance the network bisection bandwidth, instead of using expensive high speed routers. Our protocol DiFS is designed for an arbitrary multi-rooted tree topology. However for the ease of exposition and comparison with existing protocols, we will use the fat-tree topology for our protocol description and experimental evaluation.

\begin{figure}[t]
\centering
\includegraphics[width=9cm]{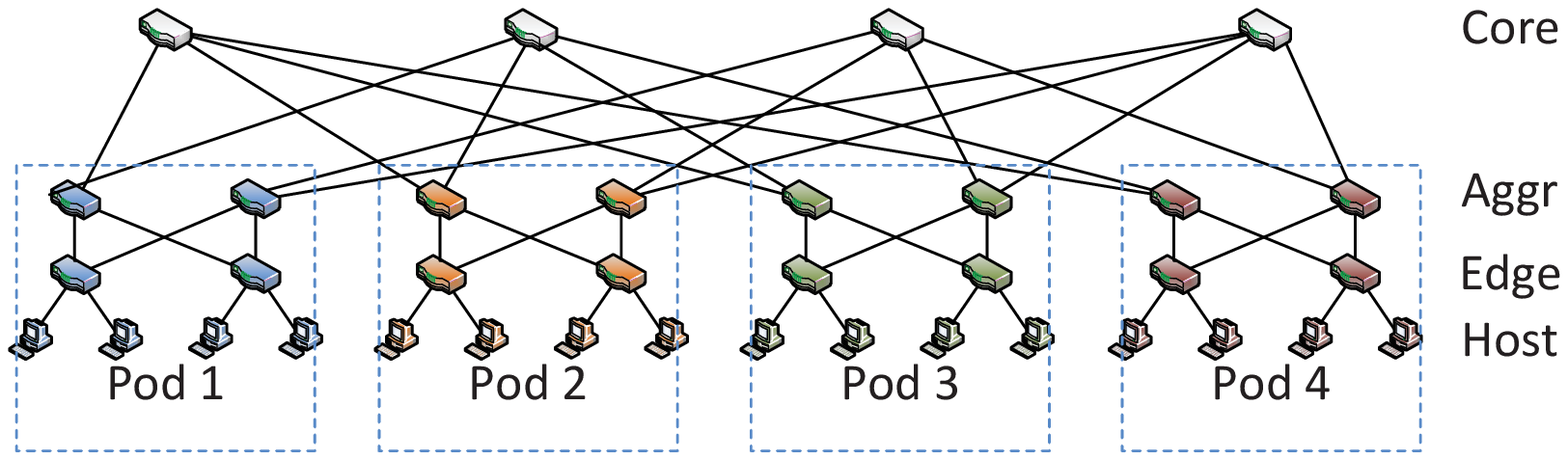}
\caption{A fat tree topology for a datacenter network}\label{fig:topology}
\vspace{-4ex}
\end{figure}

A multi-rooted tree topology has three vertical layers: edge layer, aggregate layer, and core layer. A \emph{pod} is a management unit down from the core layer, which consists of a set of interconnected end hosts and a set of edge and aggregate switches that connect these hosts.
As illustrated in Figure \ref{fig:topology}, a fat-tree network is built from a large number of $k$-port switches and end hosts.
There are $k$ pods, interconnected by $(k/2)^2$ core switches. Every pod consists of $(k/2)$ edge switches and $(k/2)$ aggregate switches. Each edge switch also connects $(k/2)$ end hosts. In the example of Figure \ref{fig:topology}, $k=4$.

A path are a set of links that connect two end hosts.
There are two kinds of paths in a fat-tree network: inter-pod path and intra-pod path.  An intra-pod path interconnects two hosts within the same pod while an inter-pod path is a path which connects two end host in different pods. Between any pair of end hosts in different pods, there are $(k/2)^2$ equal-cost paths, each of which corresponding to a core switch.  An end-to-end path can be split into two flow segments\cite{NIRA}: the \emph{uphill segment} refers to the part of the path connecting source host to the switch in the highest layer (e.g., the core switch for an inter-pod path), and the \emph{downhill segment} refers to the part connecting the switch in the highest layer to the destination host. Similar to existing work, we mainly focus on flows that take inter-pod paths, because they cause most flow collisions.


\begin{figure*}[t]
\begin{center}
\begin{tabular}{p{125pt}p{160pt}p{180pt}}
\subfigure[Local collision]
 {
   \includegraphics[width=4cm]{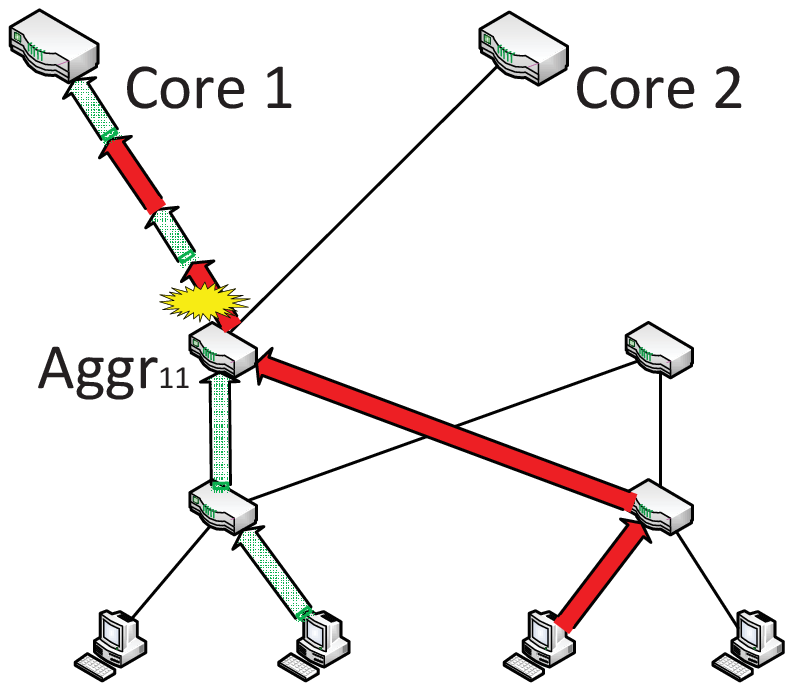}
   \label{fig:local}
 }
 &
\subfigure[Remote collision Type 1]
 {
   \includegraphics[width=5.5cm]{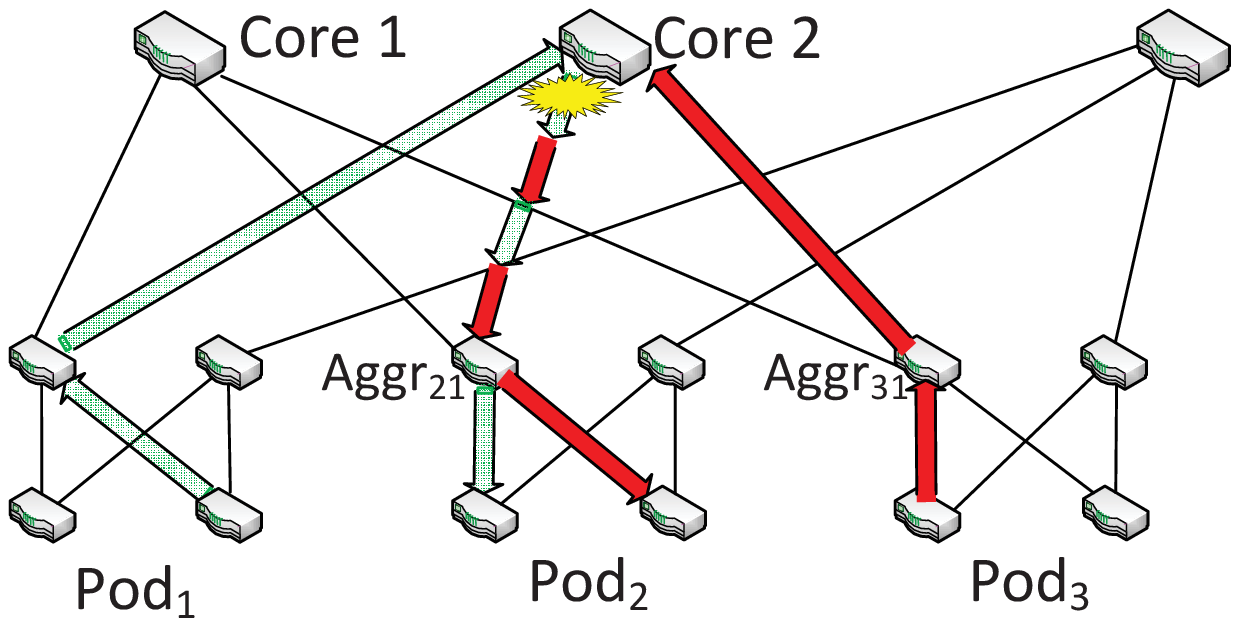}
   \label{fig:remote1}
 }
 &
\subfigure[Remote collision Type 2]
 {
   \includegraphics[width=6.5cm]{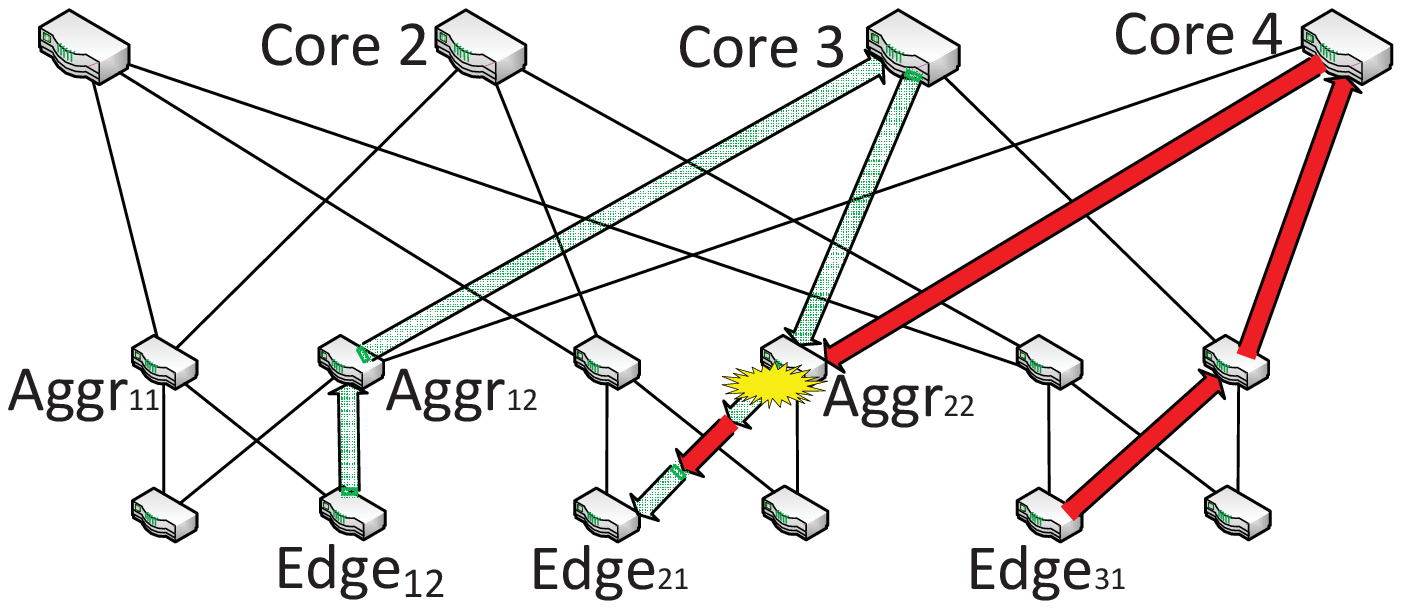}
\label{fig:remote2}
 }
\end{tabular}
\vspace{-1ex} \caption{Three types of collisions} \label{fig:collision}
\end{center}
\vspace{-4ex}
\end{figure*}

\begin{figure*}[t]
\begin{center}
\begin{tabular}{p{125pt}p{160pt}p{180pt}}
\subfigure[Path Allocation solves local collision]
 {
   \includegraphics[width=4cm]{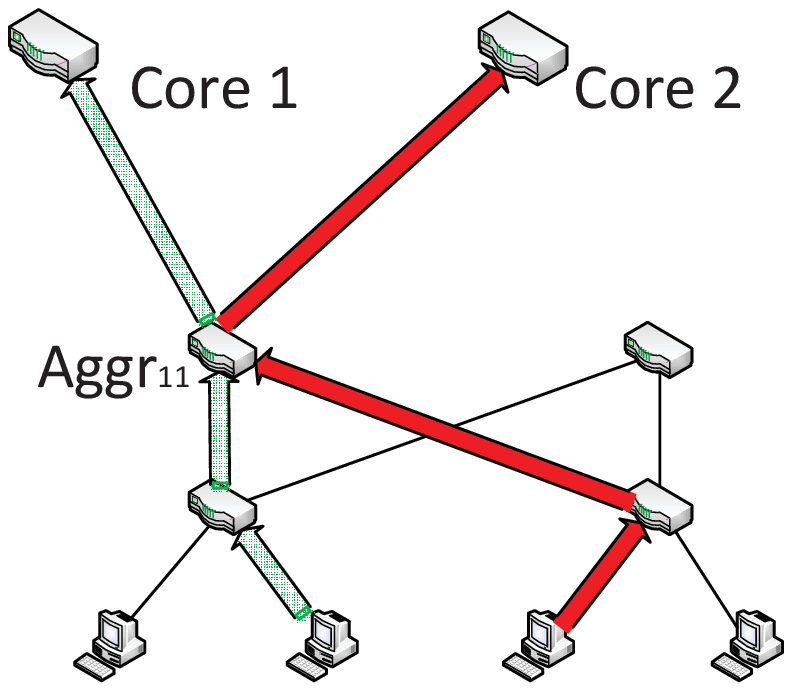}
   \label{fig:local_solve}
 }
 &
\subfigure[Solve remote collision Type 1: $Aggr_{21}$ sends back an EAR and suggest $Aggr_{31}$ to forward the flow to $Core 1$.]
 {
   \includegraphics[width=5.5cm]{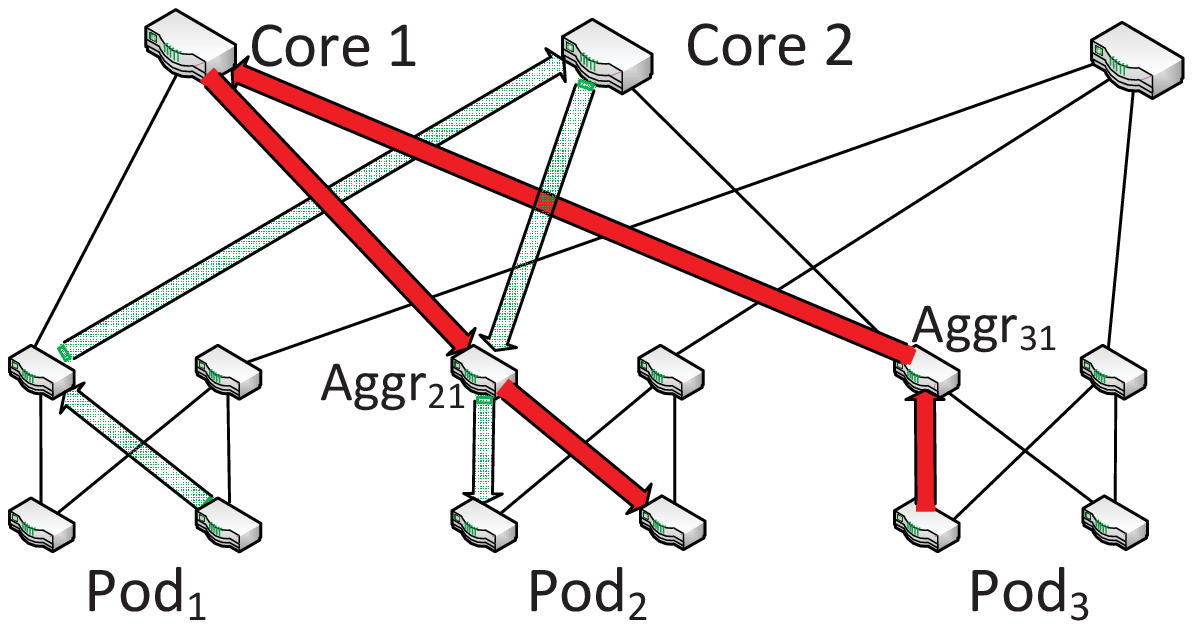}
   \label{fig:remote1_solve}
 }
 &
\subfigure[Solve remote collision Type 2: $Edge_{21}$ sends back an EAR and suggest $Edge_{12}$ to forward the flow to $Aggr{11}$.]
 {
   \includegraphics[width=6.5cm]{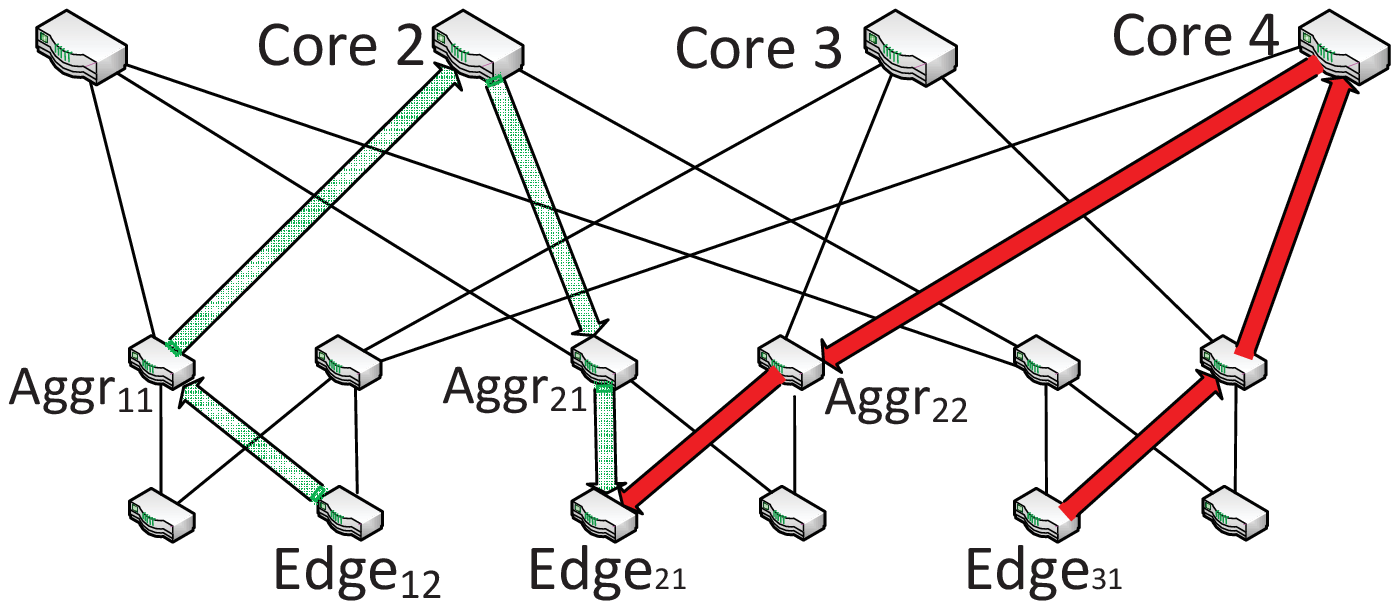}
\label{fig:remote2_solve}
 }
\end{tabular}
\vspace{-1ex} \caption{Resolving collisions by Path Allocation and Explicit Adaption} \label{fig:solve}
\end{center}
\vspace{-4ex}
\end{figure*}

\subsection{Examples of flow collision and solutions of DiFS}
We show three types of flow collisions in Figure \ref{fig:collision}, where in each example some parts of the network are not shown for simplicity. Figure \ref{fig:local} shows an example of a local collision, where switch $Aggr_{11}$ forwards two flows to a same link. Local collisions may be caused by static multipathing algorithms such as ECMP. Figure \ref{fig:remote1} shows an example of Type 1 remote collision, where two flows take a same link from $Core 2$ to $Pod_2$. Type 1 remote collision may be caused by over-utilizing a core switch (Core 2 in this example). Hence some existing work propose to balance traffic among cores \cite{Hedera}. However balancing core utilization may not be enough. Another example of remote collision (Type 2) is shown in Figure \ref{fig:remote2}, where core utilization is balanced but flows still collide the link from $Aggr_{22}$ to $Edge_{21}$.

Local collisions can be detected and resolved by local algorithms in a relatively easy way. DiFS uses the Path Allocation algorithm to detect flow-to-link imbalance and remove one of the flows to an under-utilized link, as shown in Figure \ref{fig:local_solve}. The key insight of DiFS to resolve remote collisions is to allow the switch in the downhill segment that detected flow imbalance to send an Explicit Adaption Request (EAR) message to the uphill segment. For the example of Figure \ref{fig:remote1}, $Aggr_{21}$ can detect flow imbalance among the incoming links. It then sends an EAR to $Aggr_{31}$ in $Pod_3$ (randomly chosen between two sending pods), suggesting the flow to take the path through $Core 1$. $Aggr_{31}$ runs the Explicit Adaption algorithm and changes the flow path. That flow will eventually take another incoming link of $Aggr_{21}$ as shown in Figure \ref{fig:remote1_solve}. To resolve the collision in Figure \ref{fig:remote2}, $Edge_{21}$ that detects flow imbalance sends back an
EAR and suggest $Edge_{12}$ to forward the flow to $Aggr_{11}$. That flow will eventually go from $Aggr_{21}$ to $Edge_{21}$, as shown in Figure \ref{fig:remote2_solve}.

From the examples, the key observation is that \emph{the incoming links of the aggregate (edge) switch in the downhill segment have one-to-one correspondence to the outgoing links of the aggregate (edge) switch in the uphill segment} in a multi-rooted tree. Therefore when an aggregate (edge) switch in the downhill segment detects imbalance and finds an under-utilized link, it can suggest the aggregate (edge) switch in the uphill segment to change the path to the ``mirror'' of the under-utilized link. In the example of Type 1 remote collision, $Aggr_{21}$ controls the flow to income from $Core 1$ by suggesting $Aggr_{31}$ to forward the flow to $Core 1$. In the example of Type 2 remote collision, $Edge_{21}$ controls the flow to income from $Aggr_{21}$ by suggesting $Edge_{12}$ to forward the flow to $Aggr_{11}$.

\subsection{Classification of flows}
In this paper, a flow is defined as a sequence of packets sent from a source host to a destination host using TCP.
In our flow scheduling protocol, a flow can have only one path at any time. Allowing a flow to use multiple paths simultaneously may cause packet reordering and hence reduce the throughput. However, a flow is allowed to take multiple paths at different times in its life cycle.

\textbf{Elephant and mice flows:} Elephants are large, long-lived flows whose traffic amount is higher than a threshold. The other flows are called mice flows. Similar to many other work \cite{Hedera, DARD}, \emph{our protocol focuses on elephant flows and intends to spread them as evenly as possible among all links}.  All mice flows will be processed by ECMP, because recent work has shown that ECMP forwarding can perform load-balancing efficiently and effectively for mice flows \cite{VL2}. Note that elephant flows do not necessarily require high demand of sending rates.

Let $f_{ab}$ be a flow whose source is $a$ and destination is $b$. A flow $f_{ab}$ may be classified into four types for a particular switch $s$ that runs DiFS:
$f_{ab}$ is a single-in-single-out (SISO) flow for switch $s$ if and only if there are only one possible incoming link of $s$ from $a$ and one possible outgoing link of $s$ to $b$. 
$f_{ab}$ is a single-in-multi-out (SIMO) flow for switch $s$ if and only if there are one incoming link of $s$ from $a$ and multiple outgoing links of $s$ to $b$. 
$f_{ab}$ is a multi-in-single-out (MISO) flow for switch $s$ if and only if there are multiple incoming links of $s$ from $a$ and one outgoing link of $s$ to $b$.  A close look at fat-tree networks reveals that all inter-pod flows are SIMO for the edge and aggregate switches on the uphill segments, and are MISO for the edge and aggregate switches on the downhill segments. All inter-pod flows for core switches are SISO.
Multi-in-multi-out (MIMO) flows may be defined similarly. However, there is no MIMO flow for any switch in a fat-tree network. MIMO flows may appear in general network topologies.





\section{DiFS Design}
\label{sec:design}

In this section, we present the optimization goals of DiFS protocol, the flow control modules at each switch, and the detailed scheduling algorithms.

\subsection{Optimization goals}
\label{sec:op_goal}

As a high-level description, DiFS intends to balance \emph{the number of elephant flows} among all links in the network
to utilize the bisection bandwidth and take the advantage of path diversity. We use the number of flows as the optimization metric instead of flow bandwidth consumption based on the following reasons:
\begin{enumerate}
\item A flow's maximum bandwidth consumption\footnote{A flow's maximum bandwidth consumption, also called as flow demand, is the rate the flow would grow to in a fully non-blocking network.} can hardly be estimated. As shown in \cite{Hedera}, a flow's current sending rate tells very little about its maximum bandwidth consumption. Hedera \cite{Hedera} uses global knowledge to perform flow bandwidth demand estimation. However, such method is not possible to be applied in distributed algorithms such as DiFS.
\item Using flow count as the metric, DiFS can achieve similar or even better performance compared with Hedera \cite{Hedera} and a variant of DiFS implementation that uses estimated bandwidth consumption as the metric. The results will be shown in Section \ref{sec:flowcount}.
\end{enumerate}


Two optimization goals for load-balancing scenarios are desired:

\textbf{Balanced Output (BO):} For an edge switch $s_e$, let $o(s_a)$ be the number of SIMO flows on an outgoing link connecting the aggregate switch $s_a$. 
BO of edge switch $s_e$ is achieved if and only if $o(s_{a1})- o(s_{a2}) \leq \delta$, for any two aggregate switches $s_{a1}$ and $s_{a2}$, where $\delta$ is a constant. Similarly we may define BO of an aggregate switch to cores. BO can be achieved by the Path allocation algorithm of DiFS with the smallest possible value of $\delta$ being 1.

\textbf{Balanced Input (BI):}
For an aggregate switch $s_a$, let $i(c)$ be the number of MISO flows on an incoming link connecting the core $c$.
BI of edge switch $s$ is achieved if and only if $i(c_1)- i(c_2) \leq \delta$, for any two cores $c_1$ and $c_2$, where $\delta$ is a constant. Similarly we may define BI of an edge switch from aggregate switches. BI can be achieved by Explicit Adaptation of DiFS with the smallest possible value of $\delta$ being 1.

BO and BI do not interfere with each other, and hence a switch can achieve them at a same time. \emph{Although BO and BI of a switch is optimization in a local view, we have proved that they provide global performance bounds of load balancing, as presented in Section \ref{sec:bounds}}.


\subsection{Architecture}
DiFS uses threshold to eliminate mice flows. Such threshold-based module can be installed on edge switches. It maintains the number of transmitted bytes of each flow. If this number is larger than a threshold value, the edge switch will label this flow as an elephant flow and mark the packet header to notify other switches on its path.

Each switch has a flow table which maintains three variables for every flow $f$: the incoming link identifier, denoted as $L_i$, the outgoing link identifier, denoted as $L_o$, and the last time this flow appeared, denoted as $t$. A switch also maintains two Port State Vectors (PSVs), $V_{i}$ and $V_{o}$. The $i$th element in vector $V_{i}$ is the number of flows coming from the $i$th incoming link. Likewise the $i$th element in vector $V_{o}$ is the number of flows forwarded to the $i$th outgoing link.

There are three flow control modules in aggregate and edge switches: control loop unit, explicit adaptation unit, and path allocation unit. Control loops are run periodically by switches. The main objectives of the control loop unit are to detect imbalance of MISO flows among incoming links and send Explicit Adaptation Request (EAR) if necessary. EAR is a notification message send along the \emph{reverse flow path} to recommend switches in the flow's sending pod to choose a different path. EAR includes a path recommendation. When a switch receives a EAR, it runs the explicit adaptation unit and changes the output link of the designated flow in the EAR to that on the recommended path, if possible. Path Allocation Request (PAR) is another message to request flow scheduling. PAR includes a flow identifier and requires switches to allocate an available link for this flow. Switches treat a packet with unseen flow identifier as a PAR. The sender needs to explicitly send a PAR only if path reservation is allowed to achieve a certain level of performance guarantee for upper-layer applications\cite{Oktopus}. For a SIMO flow, the path allocation unit will assign an outgoing port for this flow based on link utilization. Detailed algorithms for these modules will be presented in the following subsections.

\begin{algorithm} [t]
\DontPrintSemicolon
\SetAlgoLined
  $S$ = the set of all MISO flows forwarded by this switch\;
  \For{$f \in S$}{
    $L_i$ = incoming link of $f$\;
    $min$ = minimum value among elements in $V_{i}$ of $f$\;
    $\delta$ = imbalance threshold \;
    \If{$V_{i}[L_i] - min > T$ }{
      compute a path recommendation $p$\;
      send a EAR($f$, $p$) to $L_i$\;
      \textbf{Return}\;
    }
  }
\caption{Imbalance Detection in Control Loop}\label{lst:cl}
\vspace{-.5ex}
\end{algorithm}

\subsection{Control loop}
Each DiFS switch continuously runs a control loop. At each iteration, the switch executes the following:
\begin{enumerate}
\item Remove disappeared flows. A flow may disappear from a switch due to several reasons. For example, the flow may have finished transmission or taken another path. In each iteration, the switch will delete a flow if the difference between current time and its last-appeared time $t$ is larger than a threshold, which may be set to a multiple of the average round-trip time of flows.
\item Re-balance SIMO flows among all outgoing links.
Removing disappeared flows  may  cause the change of flow numbers on links. Thus flow re-balancing is necessary. The detailed re-balance algorithm is not presented due to space limitation.
\item Send an EAR if necessary. If the switch finds a MISO flow comes in a over-utilized link, the switch will recommend other switches to change the flow path by sending an EAR. In order to avoid TCP performance degrade caused by too many EARs, DiFS forces every switch to send at most one EAR at each iteration.
\end{enumerate}

\textbf{Imbalance detection and path recommendation for EAR:}
For fairness concern, at each iteration the switch will scan each MISO flows in a random order. The imbalance detection is also in a threshold basis, which is presented in Algorithm~\ref{lst:cl}.

Due to lack of global view of flow distribution, the EAR receiver should be told how to change the flow's path. There the EAR sender should include a path recommendation, \emph{which does not necessarily be a complete path}. In a fat-tree, both aggregate and edge switches are able to detect load imbalance and recommend an alternative path \emph{only based on local link status}.

For the flow collision example of  Figure~\ref{fig:remote1}, $Aggr_{21}$ will notice the load imbalance among incoming links and send an EAR to $Aggr_{31}$ (randomly selected between senders of the two collided flows). The path recommendation in this EAR is just $Core 1$. $Aggr_{31}$ will receive the EAR and change the flow to the output link connected with $Core 1$, and this flow will eventually come from another incoming link of $Aggr_{21}$ that was under-utilized, as shown in Figure~\ref{fig:remote1_solve}. For the flow collision example of  Figure~\ref{fig:remote2}, $Edge_{21}$ can detect it by comparing two  incoming links and then send an EAR to $Edge_{12}$ in the uphill segment. The path recommendation here is just $Aggr_{11}$. When $Edge_{12}$ let the flow take $Aggr_{11}$, the flow will eventually take another incoming link to $Edge_{21}$ and hence resolves the collision as shown in Figure \ref{fig:remote2_solve}.

As a matter of fact, in a fat-tree network a path recommendation can be specified by either a recommended core or a recommended aggregate switch in the uphill segment. For other topologies, more detailed path specification might be needed.


\begin{algorithm} [t]
\DontPrintSemicolon
\SetAlgoLined
\KwIn{Path Allocation Request $PAR$}
\KwOut{None}
  $f$ = flow identifier in $PAR$\;
  $S$ = set of links that can reach $f$'s destination\;
  \eIf{$\left\vert S \right\vert > 1$}{
    $min$ = minimal value among all $V_o[l]$, $l\in S$\;
    \For{$l \in S$}{
      \If{$V_{o}[l] > min$}{
        $S = S - \{l\}$\;
      }
    }
    $L_o$ = a random element in $S$\;
    increase $V_{o}[L_o]$ by 1\;
  }{
    $L_o$ = the first element of $S$\;
  }
  record the incoming link $L_i$ of $f$\;
  record the outgoing link $L_o$ of $f$\;
  update the access time $t$ of $f$\;
\caption{Path Allocation}\label{lst:pa}
\end{algorithm}

\subsection{Path Allocation}
In order to keep all links output balanced, we use a distributed greedy algorithm to select an outgoing link for each. When a switch received a PAR, it first check how many outgoing links can lead to the destination. If there is only one link, then the switch will simply use this link. If there are multiple links to which this flow can be forwarded, the switch will select an local optimal link for this flow. The algorithm first find the set of links with the minimum number of outgoing flows. If there are more than one links in this set, the algorithm will randomly select a link from the set.


\begin{algorithm} [t]
\DontPrintSemicolon
\SetAlgoLined
\KwIn{Explicit Adaptation Request $EAR$}
\KwOut{None}
  $f$ = flow identifier in $EAR$\;
  $r$ = recommended core or aggregate switch in $EAR$\;
  $L_i$ = current incoming link of $f$\;
  $L_o$ = current outgoing link of $f$\;
  \eIf{$r$ and $s$ are connected \textbf{and} sending $f$ to $r$ can lead to the destination of $f$}{
    $L$ = the outgoing link connecting $r$\;
    \uIf{$V_{o}[L] >= V_{o}[L_o]$}{
      move a flow currently on $L$ to $L_o$\;
    }
    move $f$ to the outgoing link $L$\;
    update the link variables of changed links\;
  }{
    forward EAR to $L_i$\;
  }
\caption{Explicit Adaptation of switch $s$}\label{lst:ea}
\end{algorithm}

\subsection{Explicit Adaptation}
An EAR includes a flow identifier and a path recommendation. As mentioned, for a fat-tree network a path recommendation can be specified by either a recommended core or a recommended uphill aggregate switch. When a switch received an EAR, it  first checks if it can move the requested flow $f$ to the recommended core or aggregate switch. If not, it will forward this EAR further towards the reverse path of $f$. If moving $f$ will cause imbalance among outgoing links, the switch swaps $f$ with another flow on the recommended link. The complete algorithm is described in Algorithm~\ref{lst:ea}.

\begin{figure*}[t]
\begin{center}
\begin{tabular}{p{150pt}p{150pt}p{150pt}}
\subfigure[Stride traffic pattern]
 {
   \includegraphics[width=5.5cm]{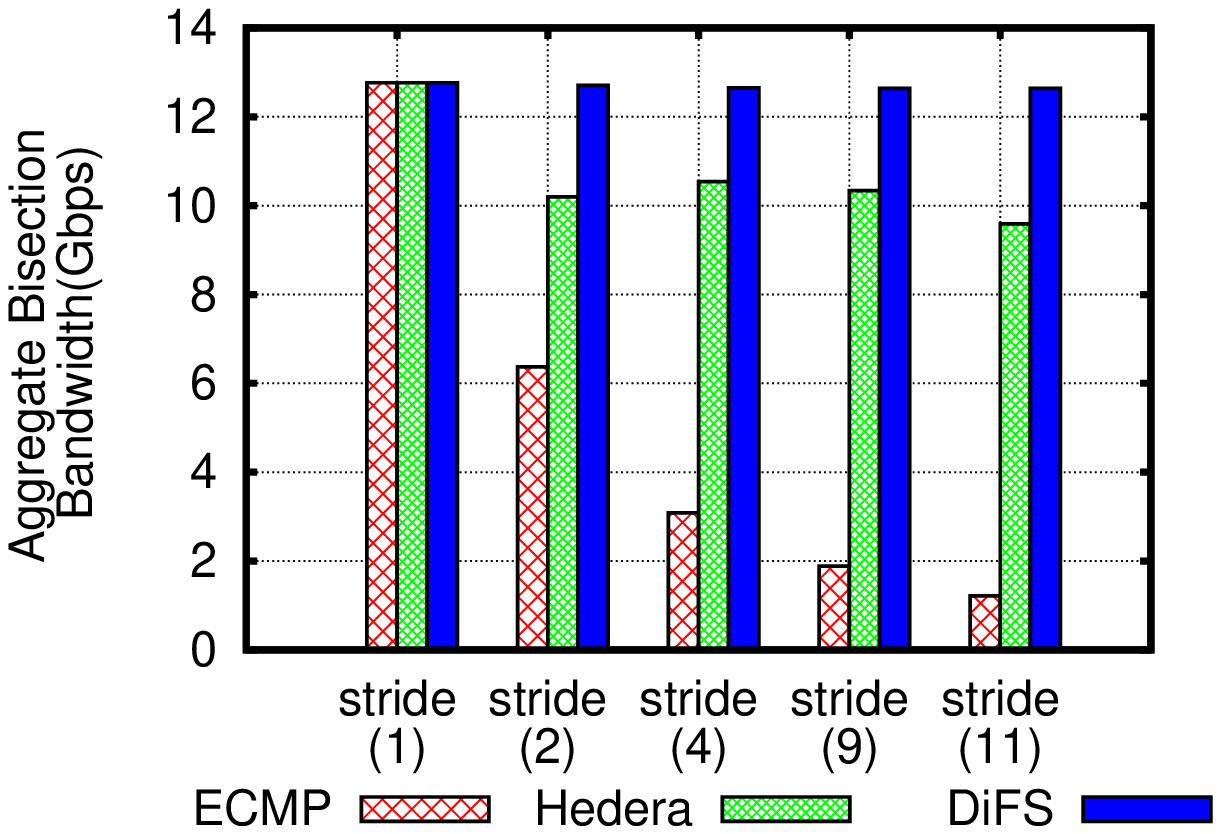}
   \label{fig:stride}
 }
 &
\subfigure[Staggered traffic pattern]
 {
   \includegraphics[width=5.5cm]{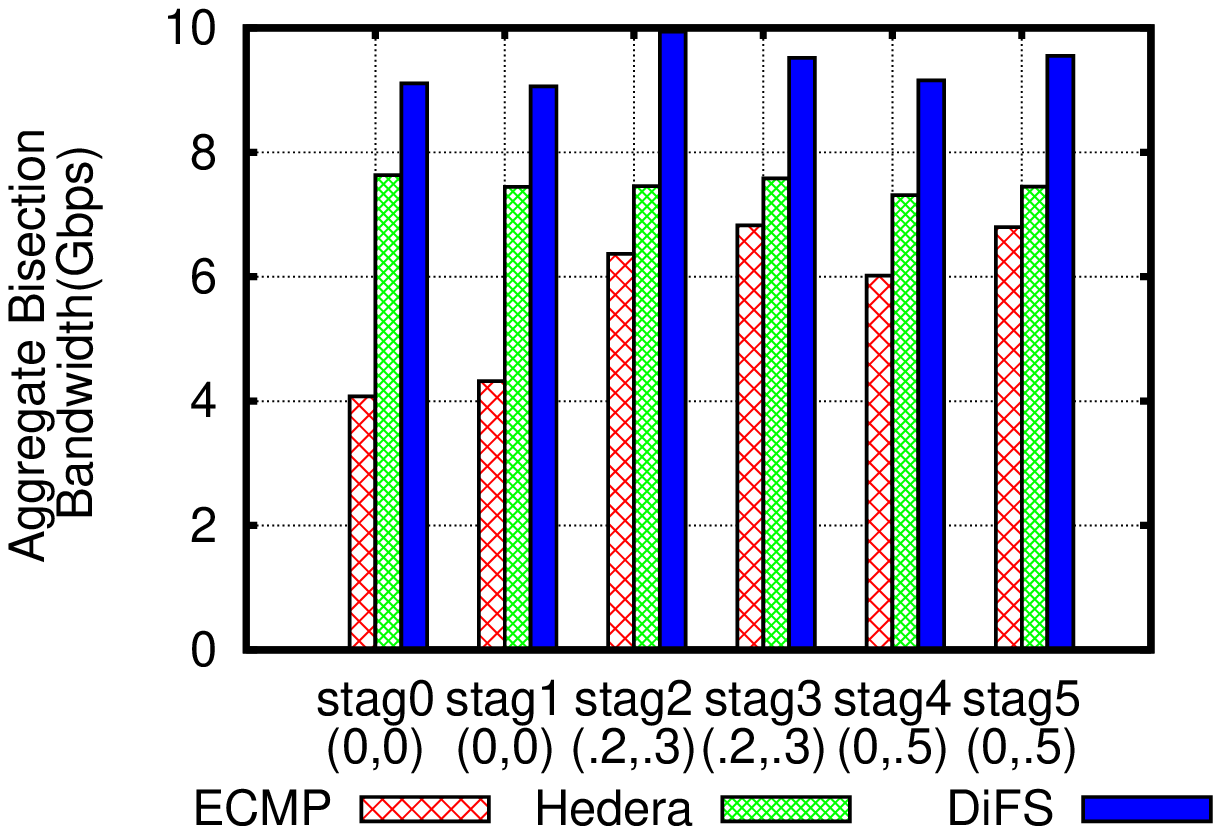}
\label{fig:stag}
 }
 &
 \subfigure[Random traffic pattern]
 {
   \includegraphics[width=5.5cm]{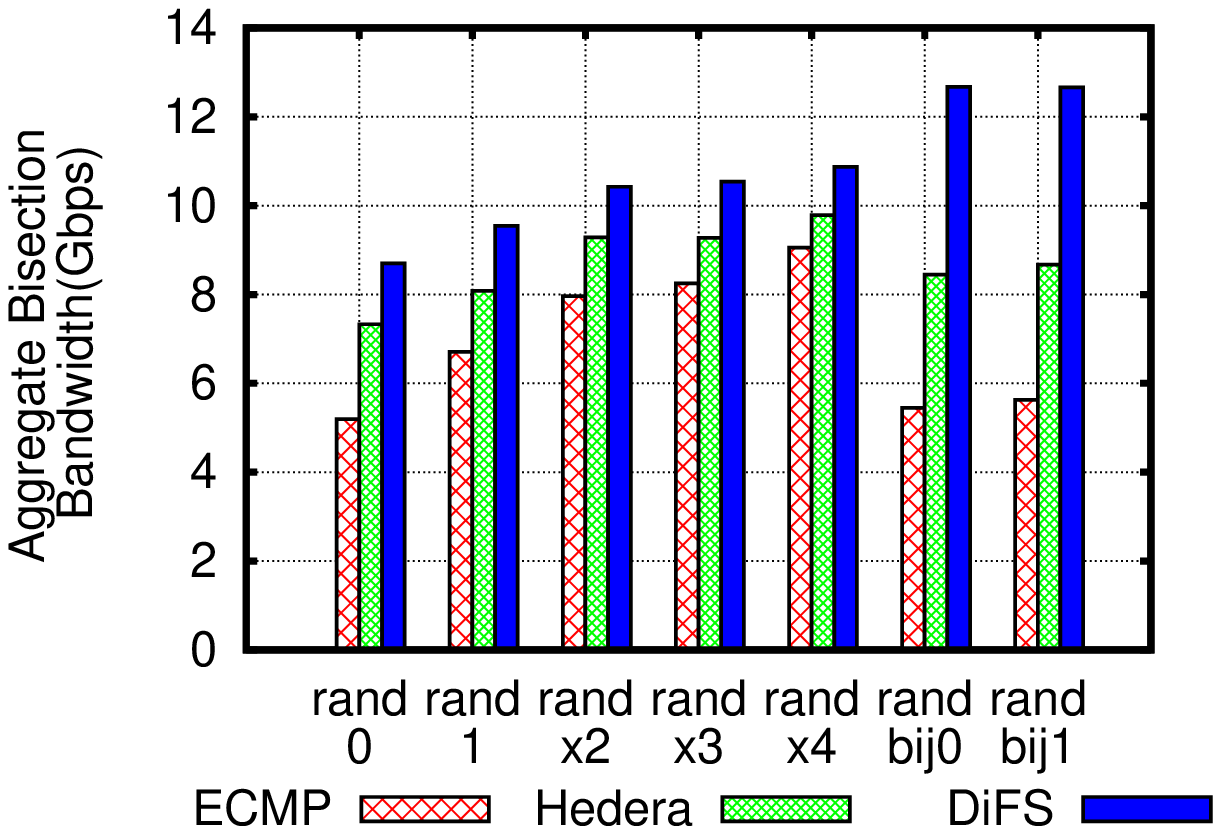}
\label{fig:rand}
 }
\end{tabular}
\vspace{-1ex} \caption{Aggregate bisection bandwidth comparison on small topologies} \label{fig:small_topo}
\end{center}
\vspace{-4ex}
\end{figure*}

\subsection{Bounds on global flow balance}
\label{sec:bounds}
The local optimization on switches can lead to global performance bounds as introduced in this section. All proofs in this section can be found in our technical report \cite{DiFS-TR}.

We provide a bound on flow balance among aggregate switches in a same pod by the following theorem:
\begin{theorem}\label{thm:bo_aggr}
In a $k$-pod fat-tree, suppose every edge switch achieves BO with $\delta$. Let $n(s_a)$ be the number of flows that are sending to aggregate switch $s_a$. Then we have $MAX_a- MIN_a \leq \delta\cdot k/2$, where $MAX_a$ is the maximum $n(s_a)$ value among all aggregate switches in the pod, $MIN_c$ is the minimum $n(s_a)$ value among all aggregate switches in the pod.
\end{theorem}

We further prove a bound on flow balance among core switches by the following theorem:
\begin{theorem}\label{thm:bo_core}
In a $k$-pod fat-tree, suppose every edge and aggregate switch achieves BO with $\delta=1$. Let $n(c)$ be the number of flows that are sending to core $c$. Then we have $MAX_{all}- MIN_{all} \leq  3k$, where $MAX_{all}$ is the maximum $n(c)$ value among all cores and $MIN_{all}$ is the minimum $n(c)$ value among all cores.
\end{theorem}

Similarly we have a bound of flow balance in the receiving side.
\begin{theorem}\label{thm:bi_edge}
In a $k$-pod fat-tree, suppose all aggregate switches in a same pod achieve BI with $\delta=1$. Let $n(s_e)$ be the number of flows that are sending to edge switch $s_e$. Then we have $MAX_{e}- MIN_{e} \leq  k/2$, where $MAX_{e}$ is the maximum $n(s_e)$ value among all edge switches in the pod and $MIN_{e}$ is the minimum $n(s_e)$ value among all edge switches in the pod.
\end{theorem}

The proof is similar to that of Theorem \ref{thm:bo_aggr}.

Note that the values we provide in the theorems are only bounds of the difference between the maximum and minimum flow numbers. In practice, however, \emph{the actual differences are much lower than these bounds}.

\subsection{Failures and Fault Tolerance}
Switches must take network failures into consideration in performing flow assignment. A network failure may be a switch failure, a link failure, or a host failure. Failures may also be classified into reachability failures and partial failures. Reachability failures refer to those failures that can cause one or more end hosts unreachable. For example, crash of an edge switch can make $(k/2)$ hosts unreachable. DiFS can tolerate such kind of failures because our algorithm relies on local, soft state collected at run time. Only flows towards the unreachable hosts are affected.

Partial failures, i.e., individual link or port failures on edge and aggregate switches, can cause performance degradation due to loss of equal-cost paths. However, DiFS can cope with such kind of failures with a simple modification. When a link or switch experiences such failure, other switches connected to the switch/link can learn the loss of capacity from underlying link state protocols.
These switches then move the flows on the failed link to other available links, or send EARs to notify the other switches.

\begin{figure*}[t]
\begin{center}
\begin{tabular}{p{150pt}p{150pt}p{150pt}}
\subfigure[Stride traffic pattern]
 {
   \includegraphics[width=5.5cm]{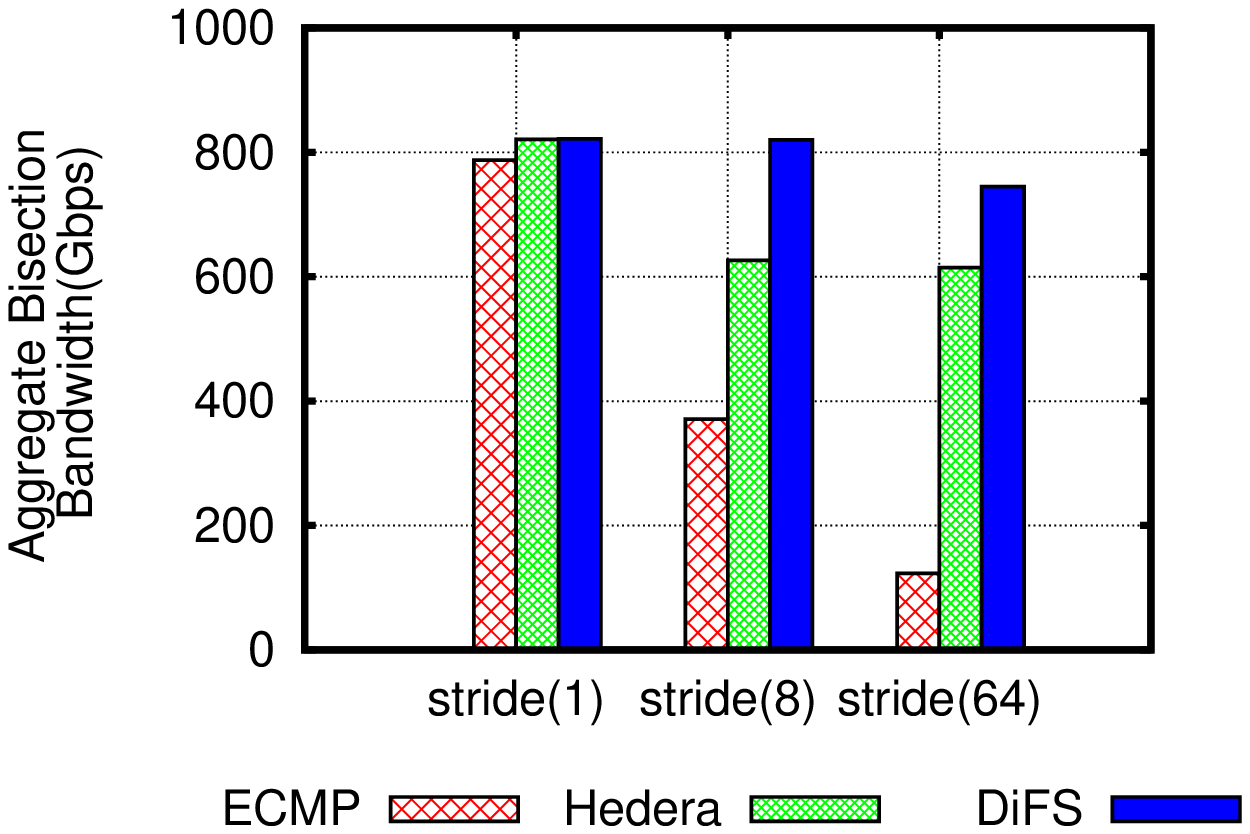}
   \label{fig:bulk1}
 }
 &
\subfigure[Staggered traffic pattern]
 {
   \includegraphics[width=5.5cm]{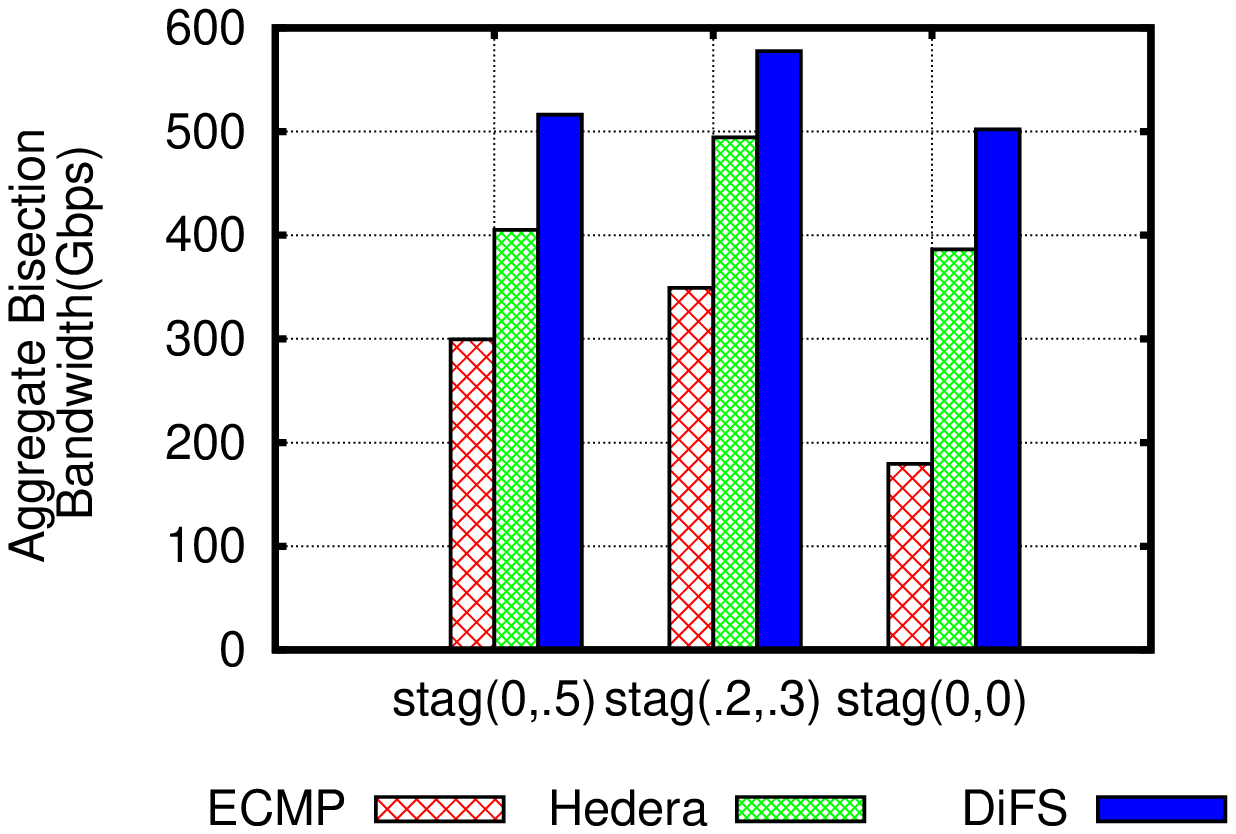}
\label{fig:bulk2}
 }
 &
 \subfigure[Random traffic pattern]
 {
   \includegraphics[width=5.5cm]{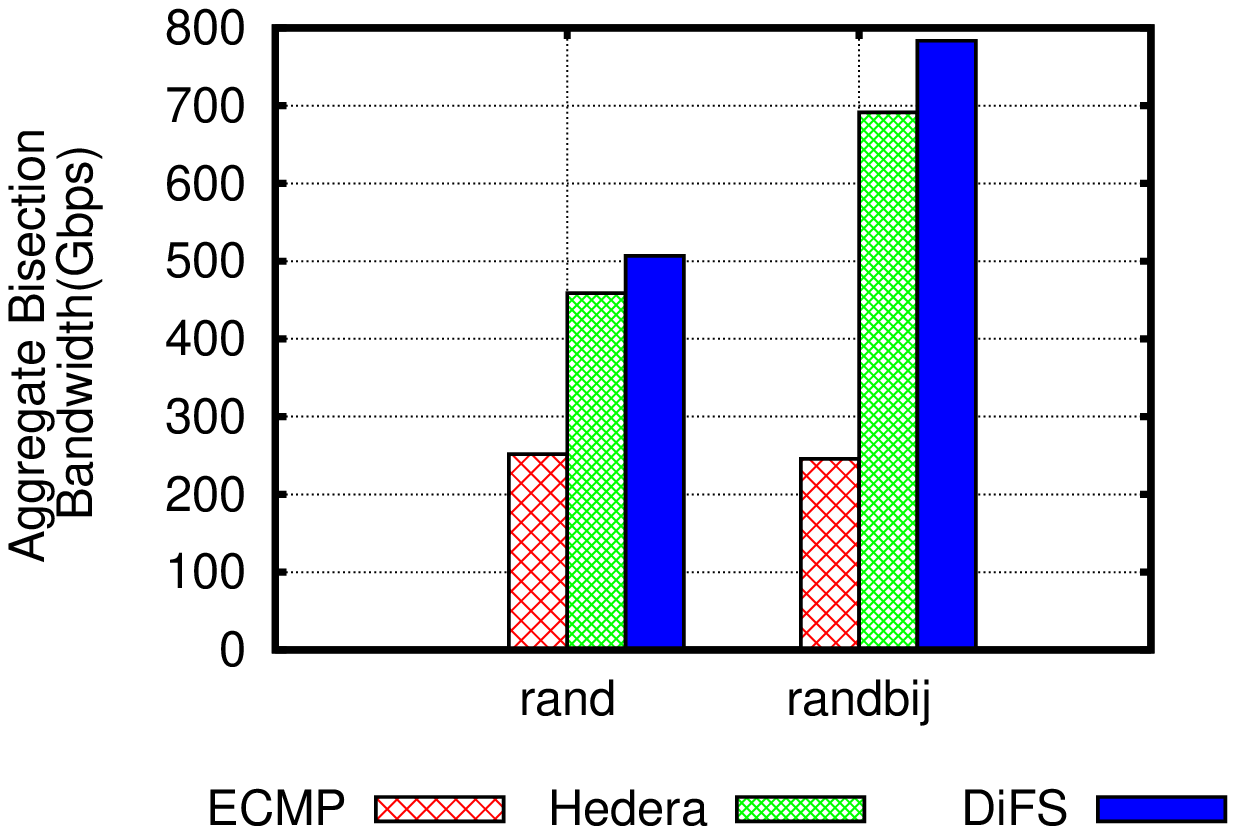}
\label{fig:bulk3}
 }
\end{tabular}
\vspace{-1ex} \caption{Aggregate bisection bandwidth comparison for bulk analysis} \label{fig:bulk}
\end{center}
\vspace{-4ex}
\end{figure*}

\section{Experimental Results}
\label{sec:simulation}
In this section, we evaluate the performance of DiFS by comparing it with two well-known solutions: a static distributed routing algorithm ECMP and a centralized scheduling algorithm Hedera \cite{Hedera}.

\subsection{Methodology}
We developed a \emph{packet-level} stand-alone simulator in which DiFS as well as other algorithms are implemented in detail. Simulation is able to show the scalability of our protocol for large networks with dynamic traffic patterns, while testbed experiments can only scale to up to tens of hosts.  The simulator developed in \cite{Hedera} only simulates each flow without performing per-packet computation, and uses predicted sending rate instead of implementing TCP. \emph{Our simulator models individual packets, hence we believe it can better demonstrate real network performance.} TCP New Reno is implemented in detail as the transportation layer protocol.
Our simulator models each link as a queue whose size is the delay-bandwidth product. A link's bandwidth is 1Gbps and its delay is 0.01ms. Our switch abstraction maintains finite shared buffers and forwarding tables.
In our experiments, we simulate multi-rooted tree topologies in different sizes. We use 16-host networks as small topologies and 1024-host networks for bulk analysis.

DiFS is compared with ECMP and Hedera. For ECMP we implemented a simple hash function which uses the flow identifier of each tcp packet as the key. We implemented the Simulated Annealing scheduler of Hedera, which achieves the best performance among all schedulers proposed in Hedera \cite{Hedera}.
We also set the period of distributed control loop to 0.01 second for DiFS. As mentioned in Section \ref{sec:op_goal}, we focus on balancing the number of elephant flows among links. We use 100KB as the elephant threshold, same to the value used by other work \cite{DARD}.

\textbf{Performance criteria.}
We focus the performance evaluation on four aspects:
\begin{enumerate}
\item
Does DiFS fully utilize bisection bandwidth? How does it compare to ECMP and Hedera?
\item
Can DiFS adapt to cluster computing applications like ``Map'' and ``Reduce''?
\item
How fast will DiFS converge to a stable state?
\item
How much is DiFS's control overhead?
\end{enumerate}



\textbf{Traffic patterns.}
Similar to \cite{Hedera} and \cite{DARD}, we created a group of traffic patterns as our benchmark communication suite. These patterns can be  either static or dynamic. For static traffic patterns, all flows are permanent. Dynamic traffic patterns refer to those in which flows start at different times. The patterns used by our experiments are described as follows:
\begin{enumerate}
\item
$Stride(i)$: A host with index $x$ sends data to a host with index $(x + i) mod (num\_hosts)$, where $num\_hosts$ is the number of all hosts in the network. This traffic pattern stresses out the links between the core and the aggregation layers with a large $i$.
\item
$Staggered(P_e, P_p)$: A host sends data to another host in the same edge layer with probability $P_e$, and to host in the same pod (but in the different edge layer) with probability $P_p$, and to hosts in different pods with probability $1 - P_e - P_p$.
\item
$Random$: A host sends one elephant flow to some other end host in the same network with a uniform probability. This is a special case of $Randx(x)$ where $x = 1$.
\item
$Randx(x)$: A host sends $x$ elephant flows to any other end host in the same topology with a uniform probability.
\item
$Randbij$: A host sends one elephant flow to some other host according to a bijective mapping of all hosts. This is a special case of $Random$ pattern which
may be created by certain cluster computing applications.
\end{enumerate}

\begin{figure}[t]
\centering
\includegraphics[width=7.5cm]{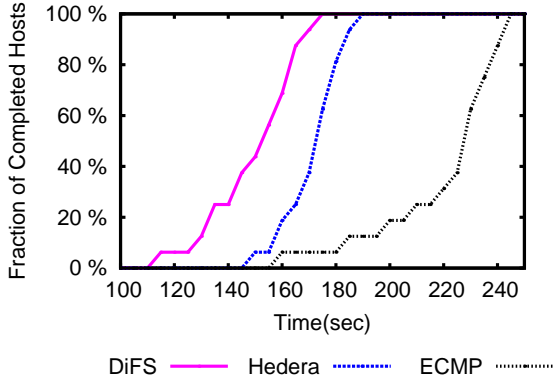}
\caption{CDF of host completion time for data shuffle}\label{fig:shuffle}
\vspace{-4ex}
\end{figure}

\subsection{Small Topology Simulation Results}

In this set of experiments, 16 hosts (acting as clients) first establish TCP connections with some designated peers (acting as servers) according to the specified traffic pattern. After that, these clients begin to send elephant flaws to their peers constantly. Each experiment last 60 seconds and each host measures the incoming bandwidth during the whole process. We use the results for the middle 40 seconds as the average bisection bandwidth.

Figure~\ref{fig:stride} shows the average bisection bandwidth for a variety of Stride traffic patterns with different parameters. For stride parameter $i=1$, all three methods have good performance. DiFS achieves highest bisection bandwidth for all $i$ values and outperforms ECMP significantly when $i$ is greater than 2.

Figure~\ref{fig:stag} shows the average bisection bandwidth for Staggered patterns. Similar to the Stride results, DiFS outperforms the others for different values of $P_e$ and $P_p$. We might find that the absolute bandwidth values of all three methods in this set of experiments are less than those in the Stride experiments. According to our results on non-blocking switches and links (not shown in the figure), the average bisection bandwidth for Staggered is also limited to 10-12 Gbps due to the hotspots created by the traffic pattern. DiFS results are actually very close to the limit.

Figure~\ref{fig:rand} depicts the bisection bandwidth for Random patterns. Random and Randomx results are similar to Staggered results, while Randombij results are similar to Stride(2) results. For all cases, DiFS outperforms Hedera and ECMP. Hedera with Simulated Annealing does not assign an explicit path for each flow. Instead Hedera assigns a core switch for every single host, which may result bottlenecks on the links connecting aggregate switches and edge switches. In Random experiments, DiFS outperforms ECMP in the average bisection bandwidth by at least $33$\% for most traffic patterns. For particular patterns, this value can be higher than $100$\%. Compared to the centralized solution Hedera,  DiFS may also achieve around $15$\% bandwidth enhancement.

\begin{figure*}[t]
\begin{center}
\begin{tabular}{p{220pt}p{220pt}}
\subfigure[16-host network]
 {
   \includegraphics[width=7.5cm]{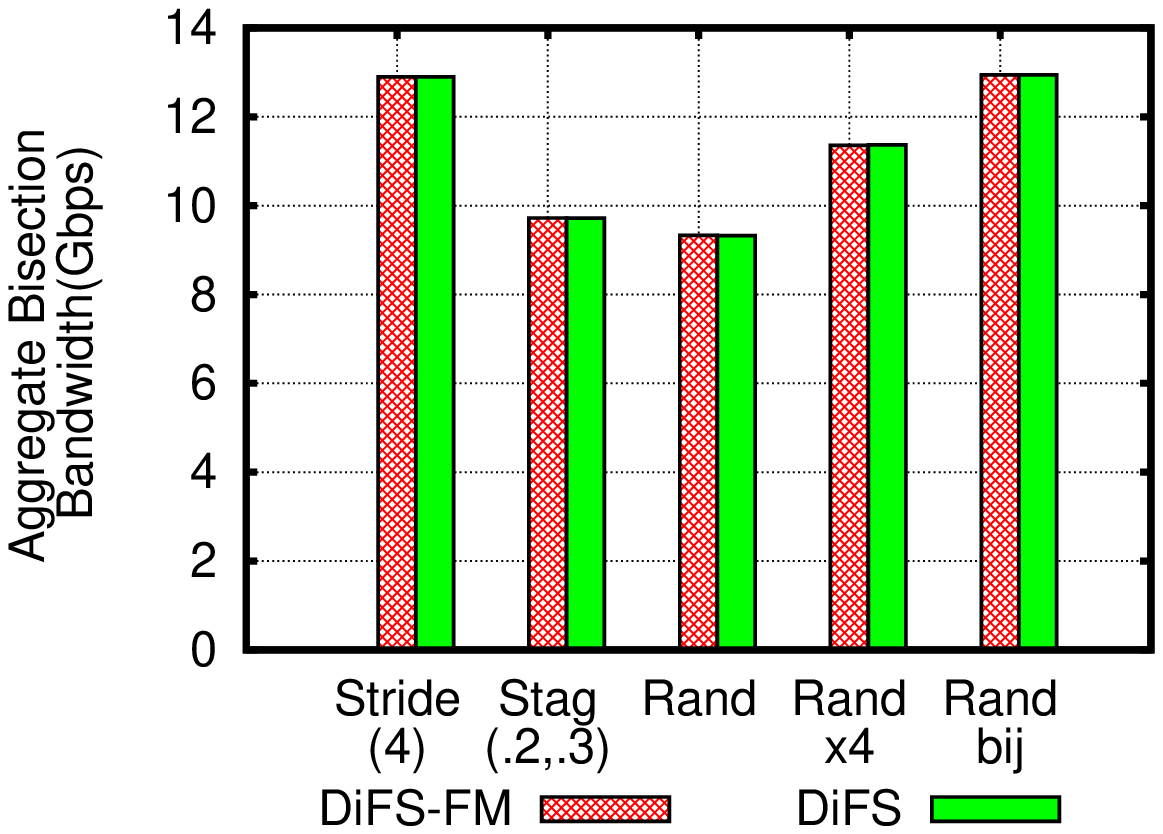}
    \label{fig:discussion1}
 }
 &
\subfigure[1024-host network]
 {
   \includegraphics[width=7.5cm]{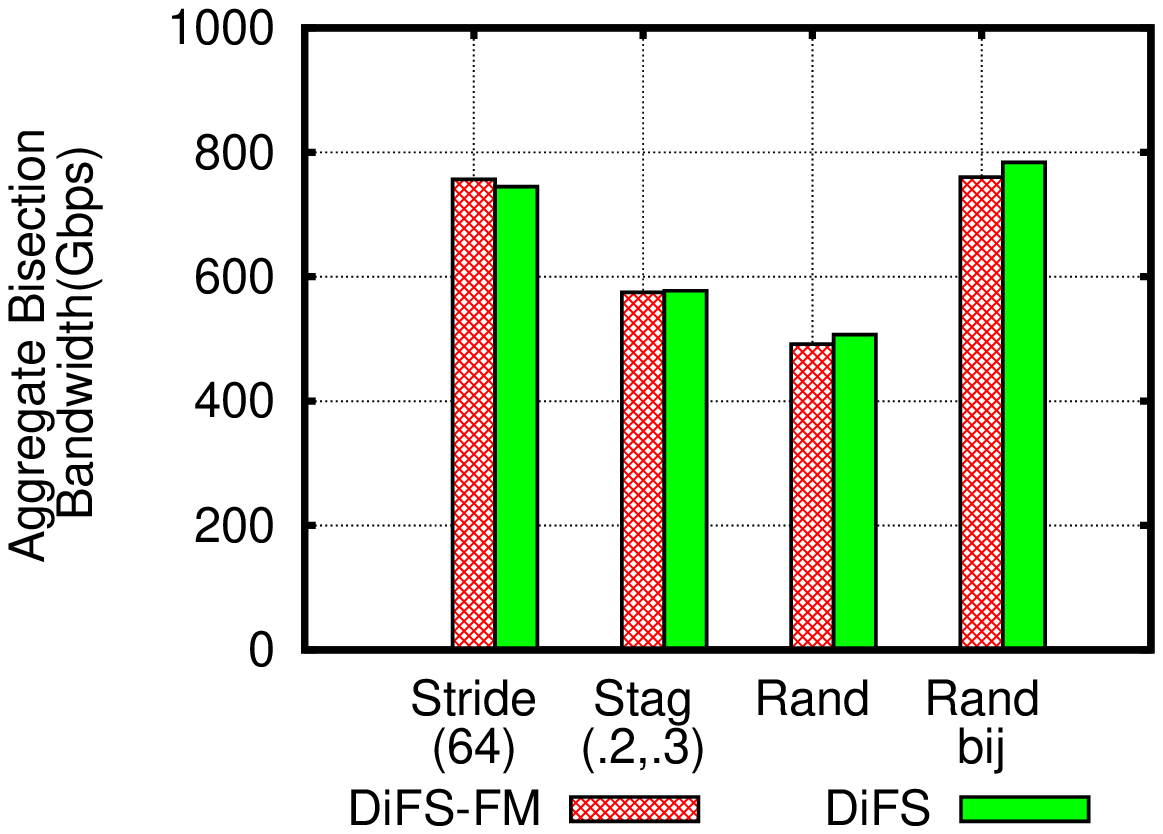}
\label{fig:discussion2}
 }
\end{tabular}
\vspace{-1ex} \caption{Flow Measuring vs Flow Counting} \label{fig:discussion}
\end{center}
\vspace{-4ex}
\end{figure*}

\subsection{Data Shuffle}
\label{sec:shuffle}
We conduct experiments of all-to-all data Shuffle in the 16-host multi-rooted tree topology to evaluate the performance of DiFS under dynamic traffic patterns. Data Shuffle is an important operation for MapReduce-like applications.  Each host (acting as reducer) in the network will sequentially receive a large amount of data (500MB in our simulation) from all other hosts (acting as mapper) using TCP. Therefore in total it is a 120GB-data Shuffle. In order to avoid unnecessary hotspot, each host will access other hosts in a random order. We also assume there is no disk operation during the whole process. We measure the shuffle time, average completion time, and average bisection bandwidth of the three methods. The shuffle time is the total time for the 120GB Shuffle operation. The average completion time is the average value of the completion time of every host in the network. The average bisection bandwidth refers to the sum of average throughput of every host.

\textbf{Packet reordering under dynamic traffic.}
We also measure two variables described in \cite{Zahavi2012} during the Shuffle period in order to reflect the packet reordering problem. The first variable is the ratio of the number of packets delivered out-of-order to the number of packets provided in-order in TCP by the senders. The second variable is the out-of-order packet window size, defined as the average gap in the packet sequence numbers observed by the receivers.

\begin{table}[t]
 \centering
 \begin{tabular}{|c|c|c|c|}
 \hline
  & ECMP & Hedera & DiFS \\ \hline  \hline
  Shuffle time (s) & 244.72 & 188.77 & 174.62 \\ \hline
  Average completion time (s) & 220.11 & 171.05 & 148.88 \\ \hline
  Average bisection BW (Gbps) & 4.29  & 5.63 & 6.53 \\ \hline
  Average out-of-order to & 0.006 & 0.009 & 0.007 \\
  in-order Ratio & & & \\ \hline
  Maximum out-of-order to & 0.009 & 0.0157 & 0.0128 \\
  in-order Ratio & & & \\ \hline
  Average out-of-order  & 0.99 & 23.32 & 13.72 \\
  window size & & & \\ \hline
  Maximum out-of-order  & 2.92 & 93.99 & 53.16 \\
  window size & & & \\ \hline
 \end{tabular}
\caption{results of shuffle experiments}\label{tab:shuffle}
\vspace{-4ex}
\end{table}

Table~\ref{tab:shuffle} shows that our algorithm outperforms ECMP by 52\% and Hedera by around 16\% in average bisection bandwidth. In addition to the average completion time shown in Table~\ref{tab:shuffle}, Figure~\ref{fig:shuffle} depicts the cumulative distribution function (CDF) of host completion time of the three methods. As observed from this figure, by the time DiFS finishes Shuffle operation, around 60\% hosts of Hedera completed their jobs and only 5\% hosts of ECMP has finished their jobs.  
All three methods have obvious variation in completion time of different hosts. Table~\ref{tab:shuffle} also shows that DiFS causes less packet reordering compared to Hedera. ECMP has the least out-of-order packets because it is a static scheduling algorithm.

\subsection{Large Topology Simulation Results}

Figure~\ref{fig:bulk} shows the aggregate bisection bandwidth comparison using a 1024-host fat-tree network ($k = 16$).
We can find that ECMP performs worse in a large topology, compared with its performance in the 16-host network using the same traffic patterns.
The performance gap between Hedera and DiFS shrinks in the 1024-host network compared to that in the 16-host network. However, DiFS still has the highest aggregate bisection bandwidth for all traffic patterns. 

\subsection{Convergence speed and control overhead}
\begin{figure}[t]
\centering
\includegraphics[width=7.5cm]{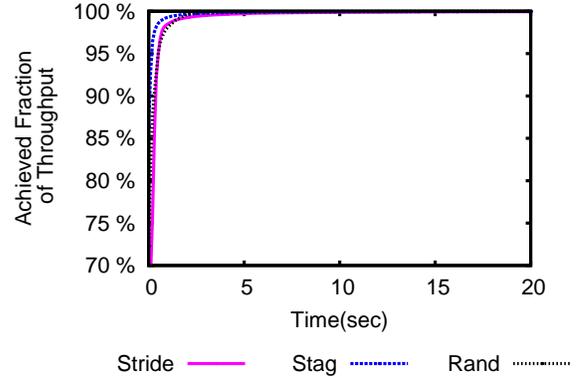}
\vspace{-2ex}
\caption{Convergence time of DiFS in the 1024-host network}\label{fig:converg}
\end{figure}

\textbf{Convergence speed.}
Convergence speed is a critical performance metric for DiFS, because DiFS is a distributed solution rather than a centralized algorithm. We measure the convergence speed of DiFS for different traffic patterns using fat-tree topologies. In Figure~\ref{fig:converg} we show the achieved fraction of throughput of DiFS versus time for different traffic patterns in the 1024-host network. Even with Random traffic our algorithm may still converge to a steady state within 5 seconds.

\begin{table}[t]
 \centering
 \begin{tabular}{|c|c|c|c|}
 \hline
  k & Host & EAR & Control Overhead(KB) \\ \hline \hline
  4 & 16 & 4 & 0 \\ \hline
  8 & 128 & 304 & 7.72 \\ \hline
  16 & 1024  & 4113 & 104.43 \\ \hline
  32 & 8192  & 45183 & 1147.22 \\ \hline
 \end{tabular}
\caption{control overhead of difs for random traffic pattern}\label{tab:overhead}
\vspace{-.5ex}
\end{table}

\begin{figure}[t]
\centering
\includegraphics[width=7.5cm]{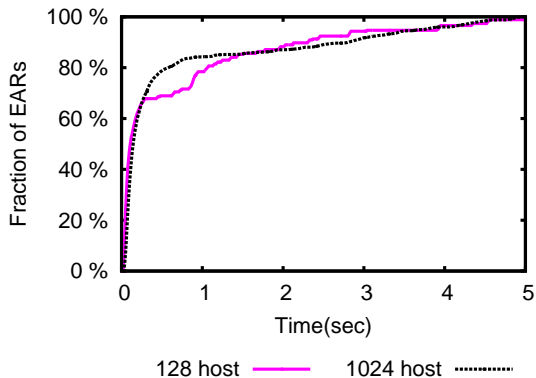}
\caption{CDF of EAR-receiving times}\label{fig:coear}
\vspace{-4ex}
\end{figure}
\textbf{Control Overhead.}
As a distributed solution, the computation cost of DiFS is very low. Hence we mainly focus on the communication overhead of DiFS, which is measured by the number of EAR messages. 
Aside from communication overhead, too many EAR messages may cause performance degradation because flows may be requested to change their paths back and forth.


Table~\ref{tab:overhead} shows the number of EARs sent by switches under random traffic patterns in fat-tree networks with different sizes. In the measurement, we assume the size of each message is 26 Bytes, which includes the size of flow identifier and the address of recommended core or aggregate switch in an EAR. As shown in the table, for an 8192-host fat-tree network, DiFS only generates control messages in a total size of around 1MB. Figure~\ref{fig:coear} shows the CDF of EAR-receiving times. Within 5 seconds, all EARs have sent and received, and around 80\% EARs are received in the first second.

\subsection{Flow count versus flow bandwidth consumption}
\label{sec:flowcount}
DiFS use the number of elephant flows as the metric for load balancing. Obviously not all elephant flows have equal bandwidth consumption, i.e., sending rate.
As discussed in Section \ref{sec:op_goal}, DiFS cannot estimate the flow bandwidth consumption due to lack of global information. A substitution for bandwidth consumption estimation is to measure the sending rate of each flow on the current path. Unfortunately, a flow's current sending rate doest not reflect its maximum bandwidth consumption \cite{Hedera}.
We also implemented a variant of DiFS which uses measured flow sending rate as the metric for load balancing, denoted as DiFS-FM. We compare both algorithms in Figure~\ref{fig:discussion1} and Figure~\ref{fig:discussion2}. The results tell that DiFS-FM has similar performance compared to DiFS that uses flow count. Therefore there is no need to deploy a particular module to keep measuring sending rates in switches.

\section{Related Works}
Recently there have been a great number of proposals for data center network topologies that provide high bisection bandwidth \cite{VL2, PortLand, Monsoon, BCube, DCell}. However, current multipathing protocols like 
Equal-Cost Multi-pathing \cite{ECMP} usually suffer from elephant flow collisions and bandwidth loss. Application layer scheduling like Orchestra \cite{Orchestra} usually focuses on higher level scheduling policies such as transfer prioritizing and ignores multipathing issues in data center networks.

Most flow scheduling solutions falls into three major categories: centralized adaptive path selection, host based multipath solution, and switch-only protocols.

Centralized adaptive selection\cite{Hedera, MicroTE} usually relies on a central controller and schedules flow path at every control interval. Aside from the additional hardware and software support for communication and computation, centralized solutions  usually face scalability problems. 
Recent research \cite{IMC09, Benson10} show that centralized solutions must employ parallelism and fast route computation heuristics to support observed traffic patterns in the data center networks.

Host-based solutions \cite{MPTCP, DARD} enable end hosts select flow path simultaneously to enhance parallelism. MPTCP \cite{MPTCP} allows a single data stream to be split across multiple paths and use congestion control algorithm to maintain load balance. Dard \cite{DARD} is similar to MPTCP while it is transparent to applications as Dard is installed below the transport layer. 
However, host-based solutions cannot scale to the size of data centers under broadcast traffic patterns. Besides, deployment of hose-based solutions requires updates on legacy systems and applications.

Switch-only protocols \cite{Kandula2005, LocalFlow, Zahavi2012} are also proposed. However most of them require flow splitting which may cause packet reordering. TeXCP \cite{Kandula2005}, as an online distributed Traffic Engineering protocols, performs packet-level load balancing by using splitting schemes like FLARE \cite{sinha2004harnessing}. Localflow \cite{LocalFlow} refines a naive link balancing solution called PacketScatter \cite{PPLB} and minimizes the number of flows that are split. \cite{Zahavi2012} also describes a general distributed adaptive routing architecture for Clos networks \cite{VL2}.

\section{conclusion}
\label{sec:conclusion}
This paper proposes DiFS, a lightweight, practical switch-only algorithms for flow scheduling in data center networks. Compared to the state-of-the-art hash-based ECMP algorithm, our algorithm can avoid bottlenecks caused by hash collision in a load-balanced manner. Instead of simply focusing on balanced output of each switch, DiFS aims to achieve balanced input by distributed switch cooperation at the same time. Simulation results also show that our algorithm can significantly outperform static hash-based ECMP and centralized scheduler like Hedera. Besides, our experiments also revealed that elephant Flow counting and bandwidth measuring have similar impacts on flow scheduling.

{\small
\bibliographystyle{abbrv}
\bibliography{pap}

\begin{thebibliography}{10}

\bibitem{PPLB}
Cisco systems. per-packet load balancing.
\newblock
  \url{http://www.cisco.com/en/US/docs/ios/12_0s/feature/guide/pplb.pdf}, 2006.

\bibitem{Al-Fares08}
M.~Al-Fares, A.~Loukissas, and A.~Vahdat.
\newblock A scalable, commodity data center network architecture.
\newblock In {\em Proceedings of ACM SIGCOMM}, 2008.

\bibitem{Hedera}
M.~Al-Fares, S.~Radhakrishnan, B.~Raghavan, N.~Huang, and A.~Vahdat.
\newblock Hedera: dynamic flow scheduling for data center networks.
\newblock In {\em Proceedings of USENIX NSDI}, 2010.

\bibitem{Oktopus}
H.~Ballani, P.~Costa, T.~Karagiannis, and A.~Rowstron.
\newblock Towards predictable datacenter networks.
\newblock In {\em Proc. of ACM SIGCOMM}, 2011.

\bibitem{Benson10}
T.~Benson, A.~Akella, and D.~A. Maltz.
\newblock Network traffic characteristics of data centers in the wild.
\newblock In {\em Proceedings of ACM IMC}, 2010.

\bibitem{MicroTE}
T.~Benson, A.~Anand, A.~Akella, and M.~Zhang.
\newblock Microte: fine grained traffic engineering for data centers.
\newblock In {\em Proceedings of CoNEXT}, 2011.

\bibitem{Orchestra}
M.~Chowdhury, M.~Zaharia, J.~Ma, M.~I. Jordan, and I.~Stoica.
\newblock Managing data transfers in computer clusters with orchestra.
\newblock In {\em Proceedings of ACM SIGCOMM}, 2011.

\bibitem{DiFS-TR}
W.~Cui and C.~Qian.
\newblock {DiFS: Distributed Flow Scheduling for Data Center Networks}.
\newblock {\em Technical Report}, 2013.

\bibitem{Dixit2011}
A.~Dixit, P.~Prakash, and R.~R. Kompella.
\newblock On the efficacy of fine-grained traffic splitting protocols in data
  center networks.
\newblock In {\em Proceedings of ACM SIGCOMM}, 2011.

\bibitem{VL2}
A.~Greenberg, J.~R. Hamilton, N.~Jain, S.~Kandula, C.~Kim, P.~Lahiri, D.~A.
  Maltz, P.~Patel, and S.~Sengupta.
\newblock Vl2: a scalable and flexible data center network.
\newblock In {\em Proceedings of ACM SIGCOMM}, 2009.

\bibitem{Monsoon}
A.~Greenberg, P.~Lahiri, D.~A. Maltz, P.~Patel, and S.~Sengupta.
\newblock Towards a next generation data center architecture: scalability and
  commoditization.
\newblock In {\em Proceedings of ACM PRESTO workshop}, 2008.

\bibitem{BCube}
C.~Guo, G.~Lu, D.~Li, H.~Wu, X.~Zhang, Y.~Shi, C.~Tian, Y.~Zhang, and S.~Lu.
\newblock Bcube: a high performance, server-centric network architecture for
  modular data centers.
\newblock In {\em Proceedings of ACM SIGCOMM}, 2009.

\bibitem{DCell}
C.~Guo, H.~Wu, K.~Tan, L.~Shi, Y.~Zhang, and S.~Lu.
\newblock Dcell: a scalable and fault-tolerant network structure for data
  centers.
\newblock In {\em Proceedings of ACM SIGCOMM}, 2008.

\bibitem{ECMP}
C.~Hopps.
\newblock Analysis of an equal-cost multi-path algorithm.
\newblock {\em RFC 2992}, 2000.

\bibitem{Kandula2005}
S.~Kandula, D.~Katabi, B.~Davie, and A.~Charny.
\newblock Walking the tightrope: responsive yet stable traffic engineering.
\newblock In {\em Proceedings of ACM SIGCOMM}, 2005.

\bibitem{IMC09}
S.~Kandula, S.~Sengupta, A.~Greenberg, P.~Patel, and R.~Chaiken.
\newblock The nature of data center traffic: measurements \& analysis.
\newblock In {\em Proceedings of ACM IMC}, 2009.

\bibitem{reorder}
K.~C. Leung, V.~Li, and D.~Yang.
\newblock An overview of packet reordering in transmission control protocol
  (tcp): Problems, solutions, and challenges.
\newblock {\em IEEE Transactions on Parallel and Distributed Systems}, 2007.

\bibitem{PortLand}
R.~Niranjan~Mysore, A.~Pamboris, N.~Farrington, N.~Huang, P.~Miri,
  S.~Radhakrishnan, V.~Subramanya, and A.~Vahdat.
\newblock Portland: a scalable fault-tolerant layer 2 data center network
  fabric.
\newblock In {\em Proceedings of ACM SIGCOMM}, 2009.

\bibitem{LocalFlow}
S.~Sen et~al.
\newblock Brief announcement: Bridging the theory-practice gap in
  multi-commodity flow routing.
\newblock In {\em Proceedings of DISC}, 2011.

\bibitem{sinha2004harnessing}
S.~Sinha, S.~Kandula, and D.~Katabi.
\newblock {Harnessing TCPs Burstiness using Flowlet Switching}.
\newblock In {\em Proceedings of ACM HotNets}, 2004.

\bibitem{MPTCP}
D.~Wischik, C.~Raiciu, A.~Greenhalgh, and M.~Handley.
\newblock Design, implementation and evaluation of congestion control for
  multipath tcp.
\newblock In {\em Proceedings of USENIX NSDI}, 2011.

\bibitem{DARD}
X.~Wu and X.~Yang.
\newblock Dard: Distributed adaptive routing for datacenter networks.
\newblock In {\em Proceedings of IEEE ICDCS}, 2012.

\bibitem{NIRA}
X.~Yang, D.~Clark, and A.~Berger.
\newblock Nira: A new inter-domain routing architecture.
\newblock {\em IEEE/ACM Transactions on Networking}, 2007.

\bibitem{Zahavi2012}
E.~Zahavi, I.~Keslassy, and A.~Kolodny.
\newblock Distributed adaptive routing for big-data applications running on
  data center networks.
\newblock In {\em Proceedings of ACM/IEEE ANCS}, 2012.

\end{thebibliography}
}

\section*{Appendix}
\textbf{Proof of Theorem \ref{thm:bo_aggr}:}
\begin{proof}
Let $x$ and $y$ be arbitrary  two aggregate switches. Let $n_{ae}$ be the number of flows from edge switch $e$ to aggregate switch $a$.
\[ n(x) = \sum n_{xe}\]
\[ n(y) = \sum n_{ye}\]
Since $|n_{xe}- n_{ye}|\leq \delta$ for every edge switches $e$ and there are $k/2$ edge switches in a pod,
\[ |n(x)- n(y)| \leq \sum|n_{xe}- n_{ye}| \leq \delta\cdot k/2\]
Hence $MAX_a- MIN_a \leq \delta\cdot k/2$.
\end{proof}

\textbf{Proof of Theorem \ref{thm:bo_core}:}
\begin{proof}
We know there are $(k/2)^2$ cores. They can be divided into $k/2$ groups $g_1, g_2, ..., g_{k/2}$, each of which contains $k/2$ cores that receive flows from a same group of aggregate switches.


Suppose $x$ and $y$ are two cores. If they belong to a same group, we can prove $n_{x}- n_{y} \leq k/2 $ using a way similar to the proof of Theorem \ref{thm:bo_aggr}.

Consider that they belong to different groups. For a pod $p$, $x$ and $y$ connect to two different switches in $p$, because they are in different core groups. Let $s_{a1}$ and $s_{a2}$ denote the switches connecting to $x$ and $y$ respectively. We have
\[n(s_{a1})-n(s_{a2}) \leq k/2 \]
according to Theorem \ref{thm:bo_aggr}. Hence the average numbers of flows from $s_{a1}$ and $s_{a2}$ to each core are $\frac{n(s_{a1})}{k/2}$ and $n(s_{a2})/2$ respectively.
\[\frac{n(s_{a1})}{k/2}- \frac{n(s_{a2})}{k/2} \leq 1\]
Let $n_{pc}$  denote the number of flows from pod $p$ to core $c$. We have $n_{px}-\frac{n(s_{a1})}{k/2} \leq 1$ (BO of $s_{a1}$), and $\frac{n(s_{a2})}{k/2}-n_{py} \leq 1$ (BO of $s_{a2}$). Hence
\[n_{px}-n_{py} \leq 1+\frac{n(s_{a1})}{k/2}- \frac{n(s_{a2})}{k/2}+1 \leq 3\]
\[n_{x} - n{y} = \sum_p n_{px} - \sum_p n_{py} = \sum_p(n_{px}-n_{py}) \leq 3k\]
\end{proof}

\end{document}